# AFM-based Functional Tomography – To Mill or not to Mill, that is the Question!

Niyorjyoti Sharma[1,*], Kristina M. Holsgrove[1], James Dalzell[1], Conor J. McCluskey[1], Jilai He[2], Dennis Meier[2], Dharmalingam Prabhakaran[3], Brian J. Rodriguez[4], Raymond G.P. McQuaid[1], J. Marty Gregg[1] and Amit Kumar[1,*]

*Corresponding author

Email: nsharma05@qub.ac.uk and a.kumar@qub.ac.uk

[1]Centre for Quantum Materials and Technologies, School of Mathematics and Physics, Queen's University Belfast, Belfast BT7 1NN, UK

[2]Department of Materials Science and Engineering, Norwegian University of Science and Technology (NTNU), Trondheim, 7491 Norway

[3]Department of Physics, Clarendon Laboratory, Parks Road, Oxford, OX1 3PU, UK

[4]School of Physics, University College Dublin, Belfield, Dublin 4, Ireland

**Abstract**:

The electrical response of ferroelectric domain walls is often influenced by their geometry underneath the sample surface. Tomographic imaging in these material systems has therefore become increasingly important for its ability to correlate the surface-level functional response with subsurface domain microstructure. In this context, AFM-based tomography emerges as a compelling choice because of its simplicity, high resolution and robust contrast mechanism. However, to date, the technique has been implemented in a limited number of ferroelectric materials, typically to depths of a few hundred nanometers or on relatively soft materials, resulting in an unclear understanding of its capabilities and limitations. In this work, AFM tomography is carried out in $YbMnO_3$, mapping its complex domain microstructure up to a depth of around 1.8 μm along with its current pathways. A model is presented, describing the impact of interconnected domain walls within the network, which act as current dividers and codetermine how currents distribute. Finally, challenges such as tip-blunting and subsurface amorphisation are identified through TEM studies, and strategies to address them are also put forward. This study highlights the potential of AFM tomography and could spur interest within the ferroics community for its use in the investigation of similar material systems.

# I. Introduction

Atomic Force Microscopy (AFM) based tomography (layer-by-layer imaging) is starting to emerge as a viable means to investigate domain structures and associated functionalities in ferroelectric oxides. [1, 2] Tomographic piezo response microscopy (TPFM) is a variant of this approach which involves rastering the PFM tip while applying a significant compressive loading force, of several micronewtons, to sequentially remove thin layers from the sample surface while undertaking simultaneous or interleaved functional imaging. Its use in the study of ferroic oxides was spurred by the pioneering study led by Huey's group who undertook volumetric imaging of ferroelectric domains in $BiFeO_3$ thin films, where they used conductive diamond tips to investigate the size-dependence of ferroelectricity and coercive fields. [3] Furthermore, the same group demonstrated the applicability of this approach for evaluating nanoscale photoconduction in hybrid perovskite semiconductors, highlighting the feasibility of direct functional imaging during milling. [4, 5] More specifically, for ferroelectrics, most existing studies till recently have been limited to a maximum depth of approximately 500 nm. [6] With the emergence of higher stiffness tips (such as all-diamond tips developed by Adama Technologies with stiffnesses reaching 3000 N/m), higher stresses can be generated under tips, enabling tomography, and therefore direct access to functionality, for significantly higher depths approaching ~10 microns in oxide crystals. As a result, these stiffer tips offer the potential to provide significant advances in the understanding of domains, domain walls and associated functional behaviour in ferroelectrics. In this context, we have recently undertaken TAFM using these tips to unveil the existence of theoretically predicted domain wall 'saddle points' in relatively 'soft' ferroelectric lead germanate ($Pb_5Ge_3O_{11}$) [2] and confirmed the presence of such domain wall saddle points in another uniaxial ferroelectric, triglycine sulfate (TGS). [7] Such 'deep AFM-tomography' findings inspired us to ask the question: can we take AFM-based functional tomography to greater depths in typical ferroelectric oxides (particularly crystals or thick films) to investigate functional properties like domain-wall conduction in three dimensions? In this work, we undertake tomographic functional property imaging in hexagonal rare-earth manganites and show that it is indeed possible to reconstruct the property in 3D, albeit with some caveats and cautions.

Due to their hardness, hexagonal manganites offer a promising opportunity to test the limits of AFM-based functional tomography. In comparison to lead germanate, we found that ytterbium manganite is at least 4 to 5 times harder to mill (Supplementary Figure S1). In addition to providing an opportunity to investigate the efficacy of TAFM, hexagonal manganites are interesting ferroelectric materials from both physics and device implementation perspectives. [8-13] The uniqueness of hexagonal manganites arises from their complex domain structure, which is a byproduct of a primary structural phase transition known as 'trimerisation'.

[11, 14-17] The uniaxial nature of hexagonal manganites, coupled with their meandering domain walls, interconnecting the vertexes, leads to the presence of head-to-head (H-H, positive bound charge), tail-to-tail (T-T, negative bound charge) and neutral segments within the same domain wall. The charge state at each segment within a domain wall can be calculated from its orientation with respect to the polarisation axis.[18, 19] This relationship between the local charge state and the relative domain wall orientation at that point is evident in the conductive atomic force microscopy (c-AFM) maps acquired on a sample surface cut parallel to the polarisation axis ($P \parallel c$).[20] At low voltages, enhanced conduction is observed in the p-type domain wall (T-T) segments, whereas the n-type (H-H) wall segments exhibit reduced conduction compared to the domain medium, requiring larger voltages to become conducting.[13] It has been demonstrated that the charge state of the domain wall, as seen in the plane where the wall intersects the surface, can roughly predict the c-AFM measured variations in local conductance (compared to bulk). The interpretation of the current signals is further complicated by the fact that the domain walls meander in all three dimensions and form a complex domain wall network with varying local resistance contributions.[21-23] Because of domain wall curvature, some portions of the domain wall are expected to be highly conducting while other portions are highly resistive. Thus, it is clear that gaining a 3-dimensional understanding of the domain microstructure in these systems, and materials with curved domain walls in general, is crucial in terms of their application in future electronics since it could lead to precise modelling of the current paths in the associated domain walls.[20]

Various techniques have been employed to date for accessing the 3D domain structure and the associated domain wall conductivities in hexagonal manganites.[23, 24] Recently, destructive as well as non-destructive FIB and SEM techniques have been used to explore the third dimension of $ErMnO_3$ up to a significant depth.[25, 26] In addition, TPFM data was recorded in the same material up to a depth of 200 nm, thereby correlating the curvature of the domain walls (obtained from TPFM) to their conductivities through finite-element modelling. Moreover, it was suggested that the relative conductance of the domain walls (with respect to the domains) could decide a characteristic depth ('cut off length') for the material beyond which domain microstructural variations have a negligible effect on the c-AFM currents. While these approaches provide suggestive information towards a definitive assessment of 3D conduction pathways, layer-by-layer c-AFM mapping still offers one of the more definitive and direct ways of evaluating conduction pathways in such systems that exhibit tortuosity. Despite being destructive in nature, SPM tip-based milling and consequent imaging of the functional properties (such as PFM or c-AFM) are considered more robust or direct for characterising ferroelectric domains and domain walls.

In this work, we present detailed deep-tomography studies on ytterbium manganite (YbMnO$_3$), a member of the rare-earth manganite family which exhibits anisotropic domain wall conductivity linked to polar discontinuities at the domain walls. Aside from establishing the efficacy of the AFM tomography approach for milling deep into hard ferroelectrics like hexagonal manganites, we illustrate that functional information, such as local conductivity, can be mapped sequentially at different depths to recreate the 3-dimensional (3D) conduction pathways in the depth of the crystal. Using the tomographic conduction mapping data, we also propose a simple and intuitive explanation for the observed c-AFM currents in hexagonal manganites and domain wall networks in general. We also discuss some experimental challenges associated with deep AFM-based functional tomography and suggest ways to overcome them.

## II. Materials and Methods

For this study, we have used an ytterbium manganite (YbMnO$_3$) crystal as the model system. Stoichiometric amounts of high purity (>99.99%) Yb$_2$O$_3$ and MnO$_2$ powders were mixed and calcined in air at 1,100 °C and 1,200 °C for 36 h each with intermediate grinding. The reacted powder was then formed into cylindrical feed rod shapes, 10 mm in diameter and 10 cm in length, and sintered at 1,250 °C for 24 h in air. Single crystals of YbMnO$_3$ were then grown using an optical floating-zone technique (Crystal Systems Inc). The growth was carried out at a rate of 2–3 mm h$^{-1}$ in a flow (200 cc min$^{-1}$) of Ar/O$_2$ mixed gas atmosphere with a feed and seed rod rotation at 20 r.p.m. An Asylum Research MFP-3D AFM is employed for the implementation of 'deep-functional AFM tomography' in YbMnO$_3$. This is done primarily because of two reasons. Firstly, the AFM offers operations with minimal drift and secondly, it can acquire PFM data at high speeds in resonance PFM. We utilized 350 N/m diamond tips obtained from Adama Innovations for both milling and collecting functional information along with an ORCA holder enabling both PFM and c-AFM data acquisition without holder changes. For this sample, conduction was only detected at the T-T domain walls within the investigated voltage regime (supplementary Figure S2). All image analysis is performed using MATLAB.

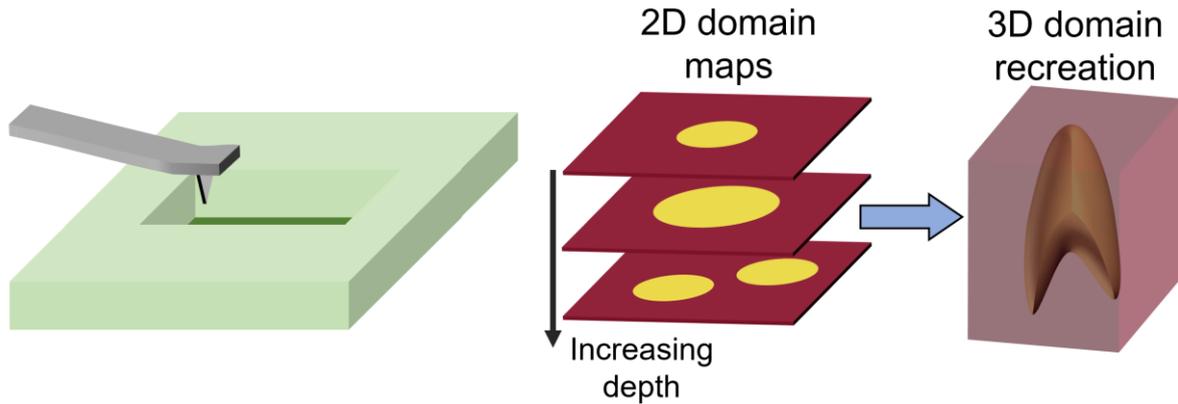

**Figure 1**: Schematic representation of AFM tip-based layer-by-layer (2D) functional tomography (piezoresponse shown here) which is then used to reconstruct the functional property in three dimensions (3D).

## II. b Milling strategy and collecting functional information

Functional AFM tomography, by definition, implies gathering functional information of a given material, while removing material layer by layer (as shown in the schematic in Figure 1). [27, 28] The acquired slice-wise information is then rendered to generate a 3-dimensional spatial map of the property of interest. Figure 1 shows a general strategy employed for tomographic PFM. An interesting point to note is whether the imaging step is concurrent with milling or is undertaken subsequent to the milling step at different milled depths. In practice, the former approach is often employed as it simplifies AFM tomography into a single-click solution. Subsequently, the voxel length along the z-axis (milling direction) is determined by dividing the final depth of the box by the total number of scans. For example, if 100 scans were required to mill 1 µm, the voxel length would be 10 nm. This approach can be considered valid for techniques like FIB milling where the energy and the flux of the bombarded ions can be held constant, or even in AFM milling when the desired depth is achieved in fewer scans (say 10 scans).

However, in the case of deep-AFM tomography, especially for hard materials, employing the same method could pose significant errors in estimating the true depth of a data slice. It is not very surprising that this assumption of removing material at a constant rate is considerably affected due to rapid tip blunting under the high friction environment during AFM milling. A systematic analysis of the significant effect of tip blunting on the rate of milling is shown in Figure 2. The data was collected on an YbMnO$_3$ single crystal when milled with a 350 N/m diamond tip. The chosen area underwent milling for 7 to 9 scans, followed by a bigger area scan at a very low tip pressure such that the depth of the milled area could be checked without inducing further milling. This process was then repeated through an automated protocol until the rate of milling fell below a desired value. Within those 7 to 9 scans, the milling rate is assumed to be constant, and the depths of the intermediate data slices are predicted through linear interpolation. As shown in Figure 2b, a set-point difference of 3 V was sufficient for a fresh tip to mill at a rate of 12 nm/scan, but the milling rate gradually dropped to zero after about 40 scans. This underscores the importance of intermittent depth checking for minimisation of depth-related errors. To address this issue, the deflection setpoints need to be adjusted once the milling rate falls below the desired rate, such that the increase in the tip radius (blunting) is corrected through increased tip force. This is aimed at keeping the stress value constant, as the rate of milling at a given stress value can be expected to be a material-dependent constant. As shown, the deflection setpoint was therefore increased from 3 V to 14.8 V at appropriate intervals. This resulted in a nearly linear increase in the depth of the milled box, as depicted in Figure 2a. Hence, we determine the global milling rate and the average height of the voxel, which we found is around 7 nm/scan. The depth mentioned here is the maximum depth determined from the line profile taken at the centre of the milled box at each stage (supplementary Figure S3).

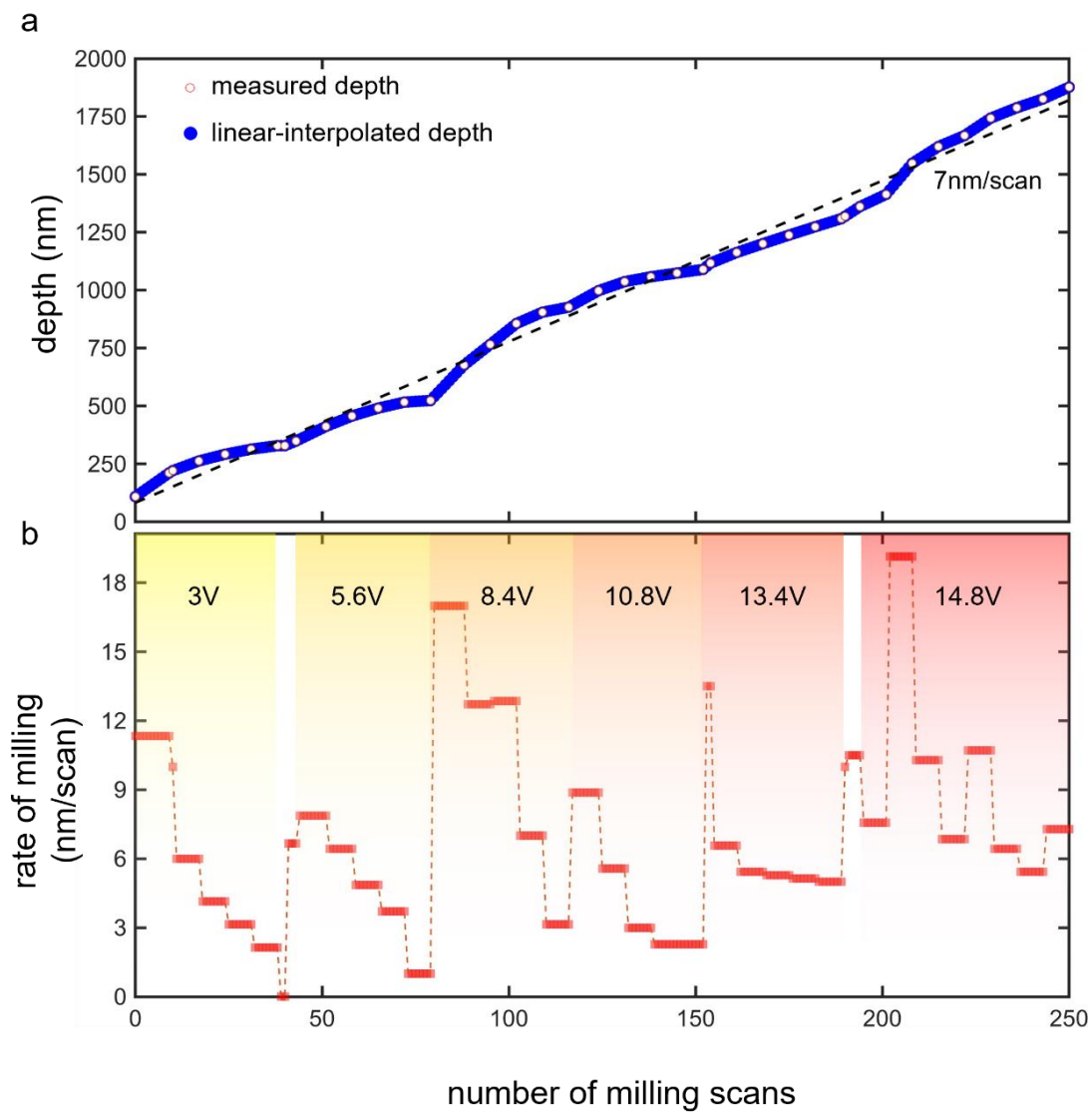

**Figure 2**: Effect of tip blunting and milling strategy (a) shows milling depth progression with increasing number of scans. The white dots represent measured depths after a specific scan number, while the blue line is the estimation for intermittent depths via linear interpolation. The dotted line represents a linear fit to the depth data (white dots). (b) depicts the rate of milling after a given number of scans at a particular tip force (in terms of deflection setpoint, V). The deflection set point was increased after appropriate intervals as shown by the colour bands. Yellow implies low force while red implies high force. A deflection set point of 1 V roughly maps to 20 µN tip force.

As mentioned, functional data in tomographic AFM can be collected in two ways, depending on the material's response and the physics under investigation, i.e., either during milling or intermittently after a few milling scans. For YbMnO$_3$, our goal was to collect PFM maps as well as undertake conduction mapping at different depths of the sample. PFM maps were obtained during milling and then c-AFM maps were collected at an interval of a few milling scans (Figure 3a). This approach was chosen because PFM provides comprehensive information about the domain structure, from which the evolution of the domains and the domain walls in the third dimension can be evaluated (Figure 3b). Additionally, the PFM data can be utilised to approximate the charge states at the surface (supplementary figure S4) at each domain wall segment in a cross-section via the relation[20, 25], thereby predicting the current paths in the bulk as shown in Figure 3c. Finally, c-AFM maps are instrumental as they provide direct supporting evidence for the charge states and thereby the conducting pathways, predicted from the PFM data (Figure 3d). It is for the same reason that obtaining c-AFM maps after an appropriate depth interval provides sufficient information in this case of AFM tomography in YbMnO$_3$. The c-AFM maps are obtained by applying a positive DC voltage to the sample base.

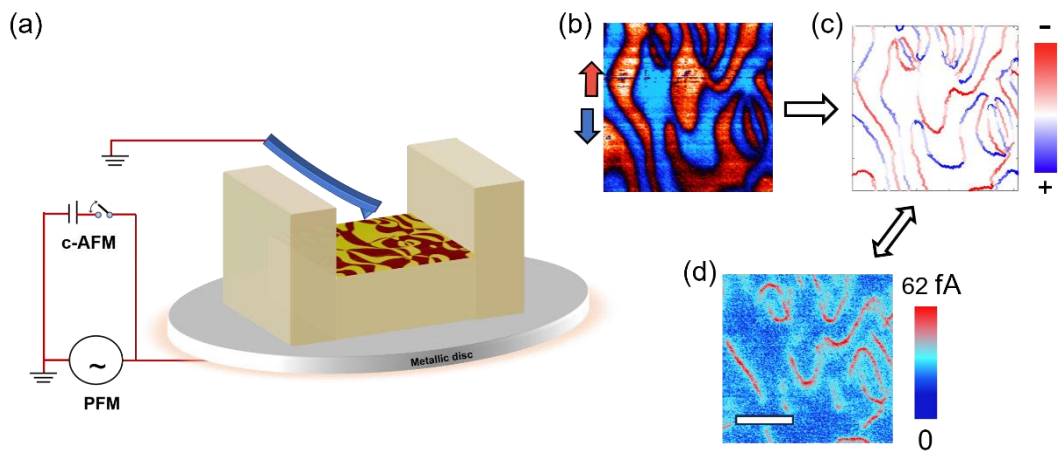

**Figure 3**: Collecting functional information during tomographic AFM: (a) Schematic illustrating how PFM and c-AFM are collected during tomographic AFM. (b) (d) PFM and c-AFM map of an area milled till the depth of 160 nm. (c) Charge state at the domain walls calculated from (b) and false coloured. +10 V was applied to the sample base during c-AFM. The scale bar in (b)-(d) is 4 µm.

## III. Results and Discussion

### IIIa. Observations from tomographic data

The PFM maps shown in Figure 4a are collected at intermittent milling depths and clearly show the evolution of the domain structure in the YbMnO$_3$ crystal. As expected, the domains follow a tortuous path through the depth of the crystal and are not isolated from one another. Domains (of the same polarity) that may appear isolated when observed on the top surface or do not appear to share the same vertex point, merge with a neighbouring domain forming new vertexes underneath the sample surface. This results in the formation of a complex domain network within the material, which has also been observed in other studies. [10, 26] With the varying domain structure in the z-direction, domain wall configurations are expected to evolve likewise. For a domain to merge with another domain of the same polarity, the domain walls must curve. As a result, the charge state of the domain walls, and thus its conductivity, would be expected to vary depending on its orientation with respect to the polarisation axis ($P \parallel c$) at a given cross-section. The variations in the charge state of domain wall segments as calculated on the newly excavated surfaces going into the depth of the material are depicted in Figure 4b along with the corresponding c-AFM maps in Figure 4c. As the domain walls separate and merge, the newly found vertexes act as a current splitter or a merging point. [21, 26] Because of these distinct characteristics in hexagonal manganites, the path associated with a current signal measured at the surface can be highly non-trivial. This effect of current spreading through the interconnected domain walls in hexagonal manganites makes them unique and can have profound implications on the currents collected during a c-AFM measurement at the top surface.

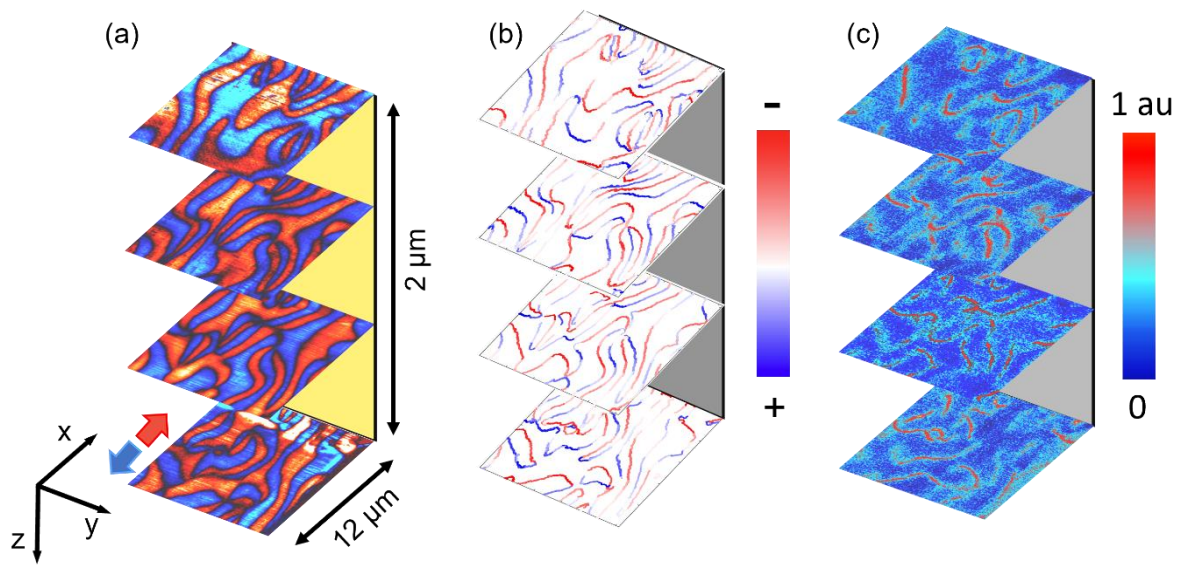

**Figure 4**: Tomographic PFM and c-AFM in YbMnO$_3$: Intermittent a) PFM b) Calculated charge state c) normalised c-AFM map of a 12 by 12-micron area in YbMnO$_3$ up to a depth of 2 μm. The slices are taken at depths of 160 nm, 1400 nm, 1800 nm and 2000 nm.

A closer look at the evolution of domain structure until a depth of 1800 nm in an adjacent area is shown in Figure 5. The PFM images at different depths are binarized, manually aligned and stacked to render a 3D domain structure revealed in Figure 5a (while keeping one of the domain variants transparent to help aid visualisation). As expected, the domains meander in a complicated fashion. Moreover, current dividers responsible for spreading a localised current signal are apparent and two of them are highlighted.

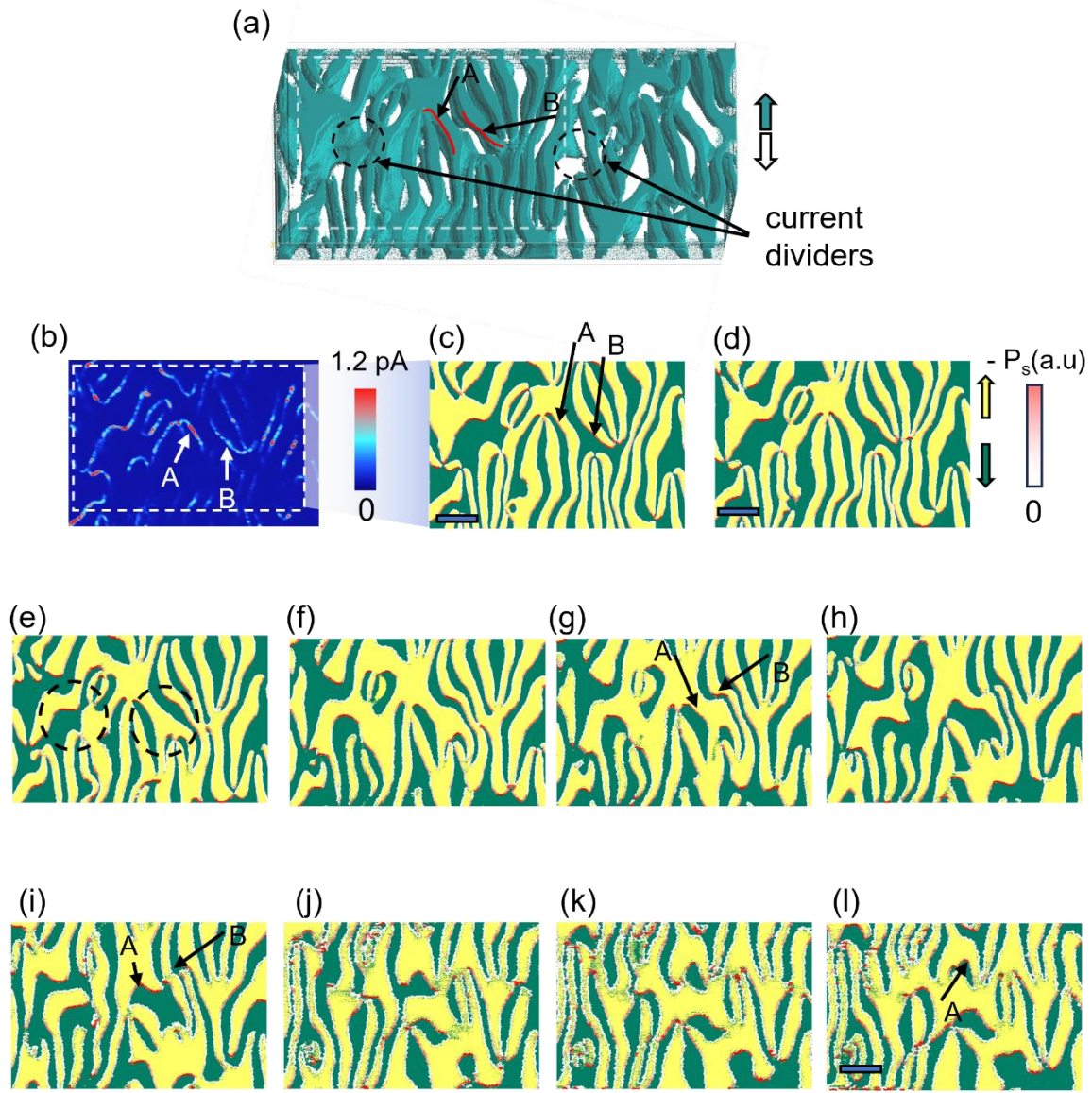

**Figure 5**: Evolution of domain structure in hexagonal manganites: (a) 3D rendering of the domain structure obtained from binarized PFM data till a depth of around 1.8 μm. Its dimension is 28 by 12.7 by 1.8 μm³. (b) c-AFM map of the area before milling. (c) to (l) are the binarized PFM maps superimposed on the calculated charge states of the T-T walls. They are collected at depths of 0, 211, 396, 660, 884, 967, 1077, 1450, 1626, and 1788 nm respectively. The white to-red colour bar corresponds to the charge state at the T-T domain walls; hence the conduction pathways can be predicted. The scale bar from (c) to (l) is 3 μm. The white dashed box in (a) (b) corresponds to the area shown in (c) to (l).

An interesting question to address via functional tomography in YbMnO$_3$ is whether subsurface domain wall geometry influences the measured currents at the surface of the crystal, similar to the FIB-SEM-based tomography work conducted by Roede *et al.* on an ErMnO$_3$ single crystal. [26] A closer examination of the c-AFM map in Figure 5b of the area prior to milling revealed a domain wall segment (T-T), labelled as A, that displayed a higher current than an adjacent wall (T-T), labelled as B. Seemingly, B should have demonstrated higher or at least equivalent current to A, given its greater inclination relative to the polar axis when viewed at the surface. The solution can be found by tracking the walls into the material's depth as depicted in Figure 5(c-i). As illustrated in Figure 5(i), domain segment A rotates perpendicular to the polarisation axis, resulting in a highly conductive state underneath it. Conversely, segment B rotates nearly parallel to the polarisation axis demonstrating the opposite behaviour. This corroborates that the charge state of the domain walls beneath the sample surface co-determines the measured surface currents. [25, 26, 29] However, due to the highly winding domain walls in this system, segment A may not remain perpendicular to the polarisation vector (and thus maintain a highly conducting state) further away from the surface. Indeed, as shown in Figure 5(j-k), segment A gradually transforms into a more resistive state by rotating towards the polar axis. We can generalise this notion to other domain segments as well, implying that if tracked to a suitable depth, the same segment may display a broad range of conductivity.

This raises the question: how can we measure currents at the top surface of a crystal that is a millimetre thick? It has been shown that in a conventional c-AFM setup, the electric field profile due to tip electrode geometry can result in a "cutoff length," which is determined by variables such as the relative conductivities between domain walls and domains, the width of the domain walls, and the tip diameter. [26] The cutoff length is defined as the distance beneath the tip at which the magnitude of the electric field decreases by 75%, ultimately suggesting an "active sample volume" under the tip where variations in conductivities primarily impact the overall current. A plot for the expected cut-off lengths (without considering possibilities of domain wall networks) at different relative domain wall conductivities and tip radius is shown in the Supplementary S5(c) using derivation undertaken by Roede *et al.*. [26] The plot shows strong dependence of the cutoff length on the relative conductivities of the domains and domain walls which are expected to differ between materials. The presence of a network of interconnected domain walls within this active region could introduce an additional key parameter (vertex density) causing further changes in the cut-off length.

If we consider the domain wall network as resistors in series, the resistor with the highest resistance has the most impact on the current in the circuit. Therefore, if the charge state of a domain wall segment at some depth within the material indicates a high resistance state, regardless of its conductivity elsewhere, the c-AFM currents for the segment are expected to be low. Notably, it is almost certain that every domain wall segment that has a conductive state on the surface can be expected to transform into a high resistance state at some depth within the material, despite showing conduction at the top surface. Perhaps this is where the uniqueness of hexagonal manganites in terms of their domain wall network comes into play. Due to the domain wall network, the current signal injected at the top surface (not milled) finds an alternative route (a less resistive path) towards the ground. While a straight wall, connecting the tip and base, operates in a series model, the interconnected domain wall network in hexagonal manganites can be more appropriately be envisaged as resistors in parallel. Again, the resistor elements referred to here are the domain wall segments straddling between two vertex points that are acting as splitters. Now, when two resistors are connected in parallel, their equivalent resistance is lower than their individual resistances. Therefore, as the locally injected current spreads across the sample volume, the voltage drop across each resistive element (domain wall segment) decreases monotonically from one junction to another. This is demonstrated through a simple 2D-circuit model of randomly interconnecting resistors as shown in Supplementary S5. Importantly, we have assumed all the resistor elements to be of the same resistance, which is certainly far from the real picture and does not allow to extract length scales. However, the model does provide a qualitative intuition on how the vertex points contribute to the flow of currents within the active sample volume.

The effect of vertex density on measured current could be particularly relevant in cases where the domain walls are several orders of magnitude higher in conductance than the bulk, in thin films with small domains and, hence, a large vertex density, or when using large tip diameters comparable to the sample thickness. In essence, the key point to underscore here is that even when the domains are completely insulating (or the conductivity ratio of domain walls to domains is infinite), the unique domain wall networks in hexagonal manganites still allow for the concept of a 'cutoff length.' This effect can be expected in all ferroelectrics that exhibit interconnected domain walls, such as irridates[30] and $BiFeO_3$.[31]

**IIIb. Challenges of tomographic conducting-AFM: The damaged layer issue**

Having established the feasibility of deep AFM tomography on hard ferroelectric ceramics such as $YbMnO_3$, we now turn our focus to a potential obstacle that may arise when gathering functional tomographic-AFM data (particularly c-AFM) and discuss possible ways to overcome the same. As mentioned earlier, we have collected c-AFM maps as a direct means to assess conductivity associated with the current pathways in $YbMnO_3$, after predicting the same from the PFM domain microstructure. It would be worth mentioning here that c-AFM is a two-probe technique and thus voltage drop across the tip-sample contact resistance or the Schottky barrier is almost unavoidable.[32] Therefore, the total current collected at the tip during c-AFM is always lower than what one would expect solely from the true conductivity of the material features: the domains and the domain walls. Since the value of the contact resistance is generally unknown, the magnitude of the measured current values does lack absolute meaning by itself. Despite that, c-AFM can still provide a comparative conductivity value and enable qualitative studies.[33] Therefore, in most cases, having adequate contrast between the domains and the domain walls is more relevant and, in fact, sufficient. A point to consider is whether the current contrast between conducting domain walls and the rest of the material changes because of milling. If so, why does this happen? And how can we address it? In the context of tomographic AFM, these questions have not been comprehensively discussed before, leading to a lack of clarity about the technique's potential to provide a definitive understanding of 3D conductivity into the material's depth.

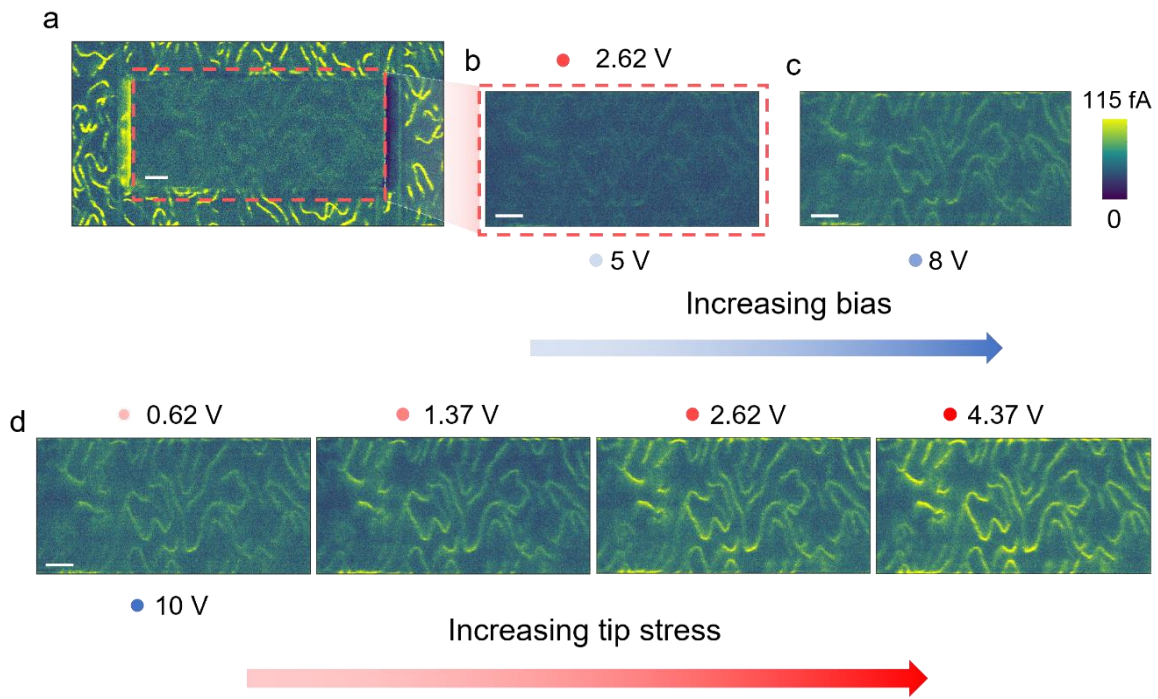

**Figure 7**: Effect of bias and tip stress on reviving currents reduced during milling: (a) c-AFM scan (5V) of a milled area and regions surrounding it. (b) and (c) are c-AFM maps obtained at the same tip stress (deflection volts: 2.62 V), however, at a sample voltage of 5V and 8V, respectively. (d) c-AFM maps of the same region were obtained at 10V sample bias but with increasing tip stress from left to right. The red circle denotes the deflection voltage (tip stress) whereas the blue dots signify the bias applied across the sample. The scale bar is 3 µm. The colour bar for all the images is the same.

To assess the effect of milling on the current in the studied sample, we obtained a c-AFM map of the area around the milled box, as depicted in Figure 7(a) (b). It is evident that the current contrast between the domains and the domain walls (T-T) has diminished within the milled region compared to the surrounding area. This milled area corresponds to the region shown in Figure 5, with an approximate depth of around 1 µm achieved in about 100 sequential scans.

Clearly, the act of milling has introduced an extra resistance to the circuit (likely beneath the tip), reducing the effective voltage drop across the active volume, ultimately leading to a poor c-AFM signal. Nevertheless, as shown in Figure 7(b)(c), the voltage across the active volume, and thereby the c-AFM contrast, can be revived to some extent by increasing the sample voltage. In addition to the sample bias, tip stress also has a similar effect on the current contrast as illustrated in Figure 7(d), possibly due to the reduction of contact resistance under high stress. [34] These experiments were conducted using a stiff 350 N/m tip; therefore, we suggest using a stiff tip at a relatively high bias to achieve good c-AFM contrast during tomographic AFM.

Although it is evident that the reduced current contrast due to milling can be attributed to an elevated resistance between the tip and the active material volume, it is imperative to understand the origin or the nature of the extra series resistance. For that, we milled a 5 by 5 by 40 nm box in a single scan, by using a 350 N/m diamond tip. The PFM and the c-AFM maps of the area before milling are shown in Figure 8 (a) and (b) respectively. As depicted in Figure 8 (c), the contrast between the domains and the domain walls (T-T) has reduced significantly post-milling. We then prepared an electron transparent lamella across the milled box, such that the c-axis (polarisation axis) of the crystal is out of plane with respect to the lamella. Transmission Electron Microscopy (TEM) revealed the presence of a subsurface damaged layer[35], depicted in Figure 8 (e), which is likely to have contributed to the extra series resistance. In regions very close to the surface of the milled box (within 10 to 20 nm) diffraction studies (5 by 5 nm area) showed an amorphous nature, as demonstrated in Figure 8 (f). For regions around 100 nm beneath the surface, crystallinity still prevails. It is worth noting that achieving 'deep'-AFM based functional tomography is trickier than milling with stiffer tips to reduce the number of scans. Typically, an optimisation procedure is needed in sync with the tomographic studies wherein a compromise is made between the depth of the damaged layer (hence c-AFM contrast or other functional properties) and the rate of milling.

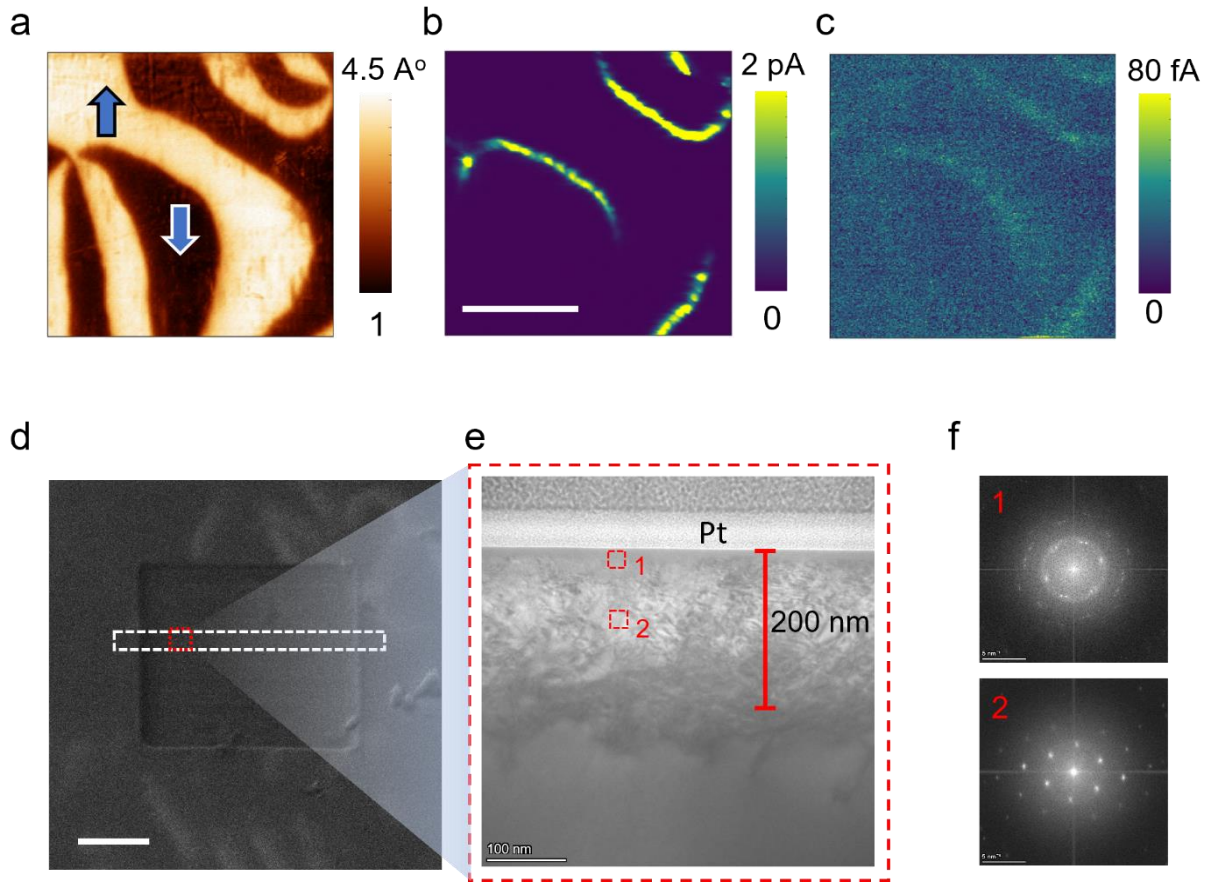

**Figure 8**: Reduced conduction due to damage-layer formation: PFM amplitude before milling (a). (b) and (c) are c-AFM maps acquired before and after milling 35-40 nm in a single scan. The sample voltage applied for (b) was 5 V, whereas it was 10 V for (c). (d) is an SEM (secondary electrons) image of the region around the milled box and the white dotted box depicts the area from where the TEM lamella was made. (e) Low-magnification TEM image of the lamella shows sub-surface damaged layer. (f) Electron diffraction pattern obtained by fast Fourier transform (FFT) of high-resolution TEM images within the boxes marked as 1 and 2 in (e). The zone axis is along [001] (c-axis). The scale bar for (a), (b), (c), and (d) is 2 µm, 100 nm for (e) and 5 nm$^{-1}$ for (f).

**IIIc. Addressing the damaged-layer issue in tomographic AFM via tip-based polishing**

This section focuses on potential solutions to address the issue of the subsurface damaged layer responsible for the reduction of current contrast during TAFM. As previously discussed, two key parameters that can be manipulated are the voltage applied to the sample and the stress on the tip. Once the existence of a damaged layer is confirmed, a significant question emerges: Is it possible to remove or at least thin the damaged layer to recover c-AFM contrast (and/or other functional property)? Theoretically, this should be achievable through 'polishing'. Mechanical polishing is a long-established technique used for reducing the roughness of a material's surface by systematically abrasing it with a surface of known lower roughness. A similar result can be obtained using an AFM tip. In this case, a chosen area is repeatedly scanned at a lower tip stress (low deflection setpoint), leading to the thinning of the damaged layer that formed after a high tip stress milling scan.

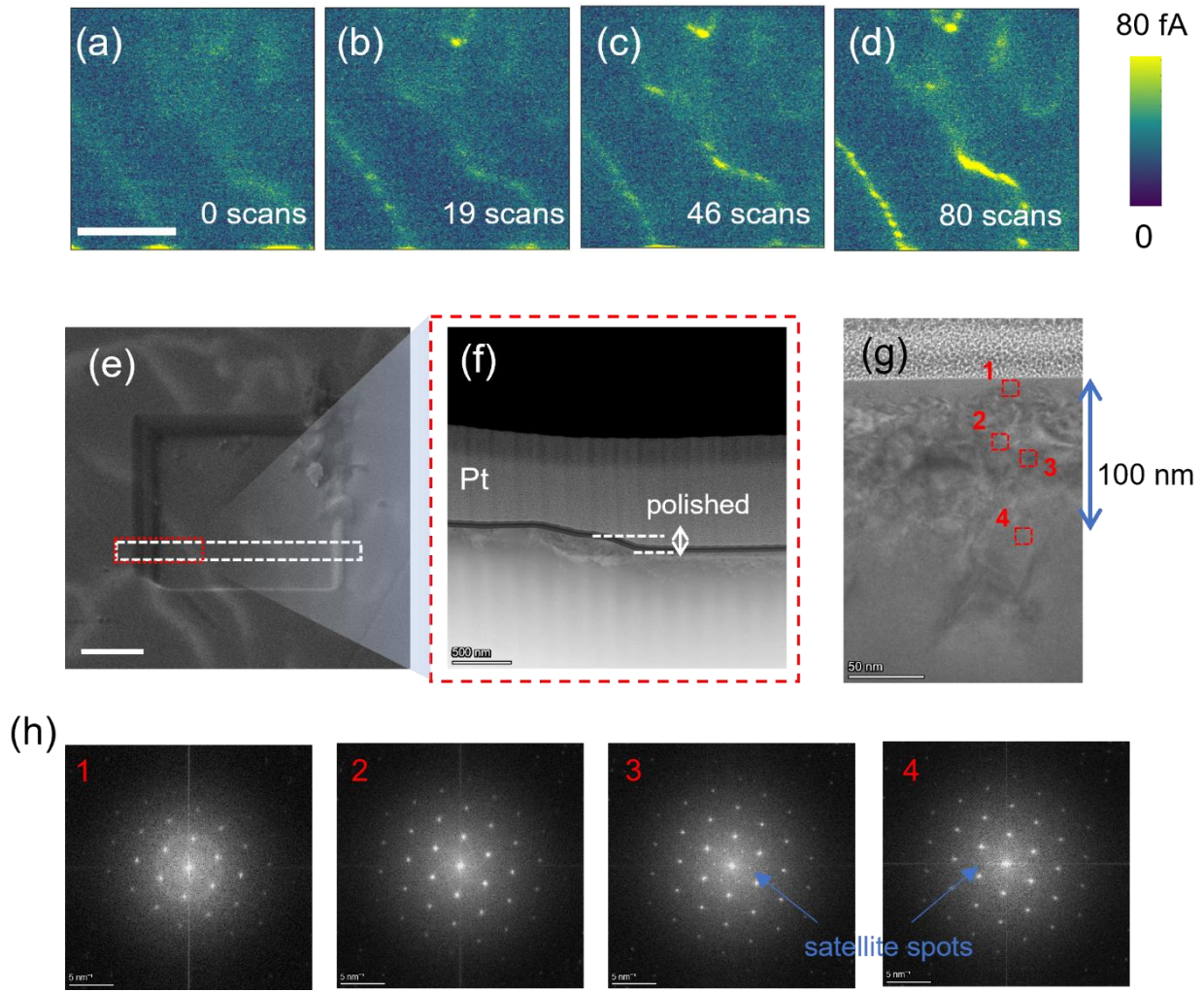

**Figure 9**: Effect of AFM polishing: (a-d) c-AFM maps acquired after 0, 19, 46, 80 polishing scans at a sample voltage of 10 V and a deflection set point of 0.58 V using a 350 N/m tip. (e) SEM (secondary electron imaging) of the milled area and its surrounding (f) HAADF-STEM image of the milled box. (g) HRTEM within the milled box. (h) FFT analysis with the red dashed boxes in (g). The scale bar in (a)-(e) is 2 µm. The scale bar for (h) is 5 nm$^{-1}$. The scale bar for (f) and (g) are 500 nm and 50 nm, respectively. The zone axis is along [001] or c-axis.

To illustrate the method, we initially milled a box measuring 5x5x40 nm$^3$ in a single scan, similar to the procedure in the preceding section. The c-AFM map obtained over the milled area after the milling scan showed a considerable decrease in current contrast. However, after conducting 'AFM polishing,' the c-AFM contrast showed a steady improvement over approximately 80 scans, as depicted in Figure 9 (a-d). During these 80 polishing scans, the deflection setpoint was kept constant. In addition to the c-AFM maps, we also compared the SEM contrast within the milled box in Figure 8(d) (unpolished) and Figure 9(e) (polished). It is evident that the conducting domain walls exhibit bright SEM contrast within the polished box, while the area inside the unpolished box is entirely dark. To confirm that this improvement is indeed due to the thinning of the damaged layer, we once again cut a lamella across the milled area for TEM analysis, as shown in Figure 9(e). The High-Angle Annular Dark-Field Scanning Transmission Electron Microscopy (HAADF-STEM) image at the left edge of the milled box (Figure 9 (f)) shows the polished region with a depth of around 150 nm. Subsequently, HRTEM was performed and diffraction analysis in 5x5 nm regions as shown in Figure 9 (g) revealed improved crystallinity compared to the unpolished case. Figure 9(h) displays the diffraction pattern at four locations at varied depths from the surface. Location 1 is closest to the surface where maximum damage is expected, while location 4 is furthest away from the surface towards the depth of the material. Locations 3 and 4 are the most crystalline, to the extent that faint satellite spots, possibly corresponding to additional ordering within the material, are also visible. As we move upwards towards the surface, the satellite spots disappear first (Location 2), and finally, near the top surface, we observe a rather subtle amorphous-like nature (Location 1), as inferred from the diffuse ring-shaped contrast in the diffraction pattern. It should be noted however, that even near the top surface (maximum damage) crystallinity has been maintained. This conclusively demonstrates the effectiveness of AFM polishing in controlling the damaged layer problem during 'deep-AFM tomography'.

## IV. Conclusion

In this work, we have demonstrated the effectiveness of the 'deep' AFM functional tomography approach in a rare-earth manganite $YbMnO_3$, which has a higher fracture point compared to other ferroelectric materials previously studied through tomographic AFM. The PFM and the c-AFM maps, collected to a depth of around 2 µm, reveal the domain wall charge states and thereby the conduction pathways in the model rare-earth hexagonal manganite. Consistent with previous reports, we find evidence of the role of sub-surface domain structure on the c-AFM currents measured on the top surface. Additionally, the influence of vertex defects, acting as current splitters or combiners within the complex domain wall resistor network, on c-AFM currents in manganites is explored using a resistor model. A current signal injected via an AFM tip can therefore be expected to split multiple times, resulting in a progressive drop in the effective circuit resistance with increasing material depth. This provides additional insight into how currents distribute within complex domain wall networks, giving a simple model to evaluate respective effects. Finally, since our tomographic approach involves high-stress applications, we identify surface amorphisation as a challenge, especially in achieving adequate tomographic c-AFM contrast, which is validated by TEM analysis. Nonetheless, additional steps such as increasing the bias, the tip pressure or thinning down the damaged layer by 'AFM tip-based polishing' are demonstrated as effective ways to regain the desired domain/domain wall contrast. We believe the tomographic AFM approach implemented in this work can be translated to other hard ferroelectric materials that are invariant under stress applications, opening up possibilities for exploring physics in the third dimension.

**Acknowledgements**: A.K. and K.H. gratefully acknowledge support from the Department of Education and Learning NI through grant USI-205 and Engineering and Physical Sciences Research Council via grant EP/S037179/1. R.McQ gratefully acknowledges support via the UKRI Future Leaders Fellowship program (MR/T043172/1). The authors gratefully acknowledge the financial support received from Tezpur University, India in the form of PhD studentship to NS under the collaborative TU-QUB PhD program. B.J.R. gratefully acknowledges support from Science Foundation Ireland via SFI/21/UUS/3765. J.H. and D.M. acknowledge funding from the European Research Council under the European Union's Horizon 2020 Research and Innovation Program (grant agreement no. 863691).

**Data availability**: The data that supports the findings of this study are available from the corresponding author upon reasonable request.

**AFM-based Functional Tomography – To Mill or not to Mill, that is the Question!**

Niyorjyoti Sharma[1, *], Kristina M. Holsgrove[1], James Dalzell[1], Conor J. McCluskey[1], Jilai He[2], Dennis Meier[2], Dharmalingam Prabhakaran[3], Brian J. Rodriguez[4], Raymond G.P. McQuaid[1], J. Marty Gregg[1] and Amit Kumar[1, *]

*Corresponding author

Email: nsharma09@qub.ac.uk  and a.kumar@qub.ac.uk

[1]Centre for Quantum Materials and Technologies, School of Mathematics and Physics, Queen's University Belfast, Belfast BT7 1NN, UK

[2]Department of Materials Science and Engineering, Norwegian University of Science and Technology (NTNU), Trondheim, 7491 Norway

[3]Department of Physics, Clarendon Laboratory, Parks Road, Oxford, OX1 3PU, UK

[4]School of Physics, University College Dublin, Belfield, Dublin 4, Ireland

1. **Comparison of hardness between Lead Germanate ($Pb_5Ge_3O_{11}$) and Ytterbium Manganite ($YbMnO_3$)**

The hardness of a material is defined by its ability to resist permanent (or plastic) deformation under a loading force.[i, ii] It is usually determined by indenting the sample with a known hard material such as a diamond probe (of a standard shape), followed by measuring the geometry (typically the projected area) of the indent. A commonly known hardness test is the 'Vickers hardness test,' where a square pyramid-shaped diamond probe is used to indent the sample, and the hardness (H) is calculated from the applied pressure (P) and projected area (A) of the square imprint given by

$$H = \frac{P}{A} \qquad \text{Eqn. 1}$$

For a softer material the projected area of the imprint is larger. Under a constant loading force, the relationship between the hardness of two materials $H_1$ and $H_2$ is given by

$$H_1 = \left(\frac{A_2}{A_1}\right) H_2 = \left(\frac{d_2}{d_1}\right)^2 H_2 \qquad \text{Eqn. 2}$$

where, $d_1$ and $d_2$ are the corresponding length of the diagonals (assuming a square indentation).

The same method can be applied on the nanometre scale by using a diamond tip in an AFM setup.[iii, iv] However, for an AFM probe, the tip and the vertical piezoelectric tube are located at opposite ends of the cantilever, causing the tip to slide on the sample surface upon indentation. As a result, elongated imprints along the cantilever are typically observed at high forces as shown in **Figure S1,** where the piezoelectric tube must lower significantly. Nonetheless, since the sliding distance is a function of force, a comparative study can still be performed between two indentations performed at the same force value. In this case, we have considered the diagonal (full-width half maximum) across the point of maximum depth for a rough estimate. Subsequently, the relationship between the hardness of YbMnO$_3$ (YMO) and Pb$_5$Ge$_3$O$_{11}$ (PGO) is calculated from equation 2 and is found to be Hardness(YMO) ~ 10 Hardness(PGO).

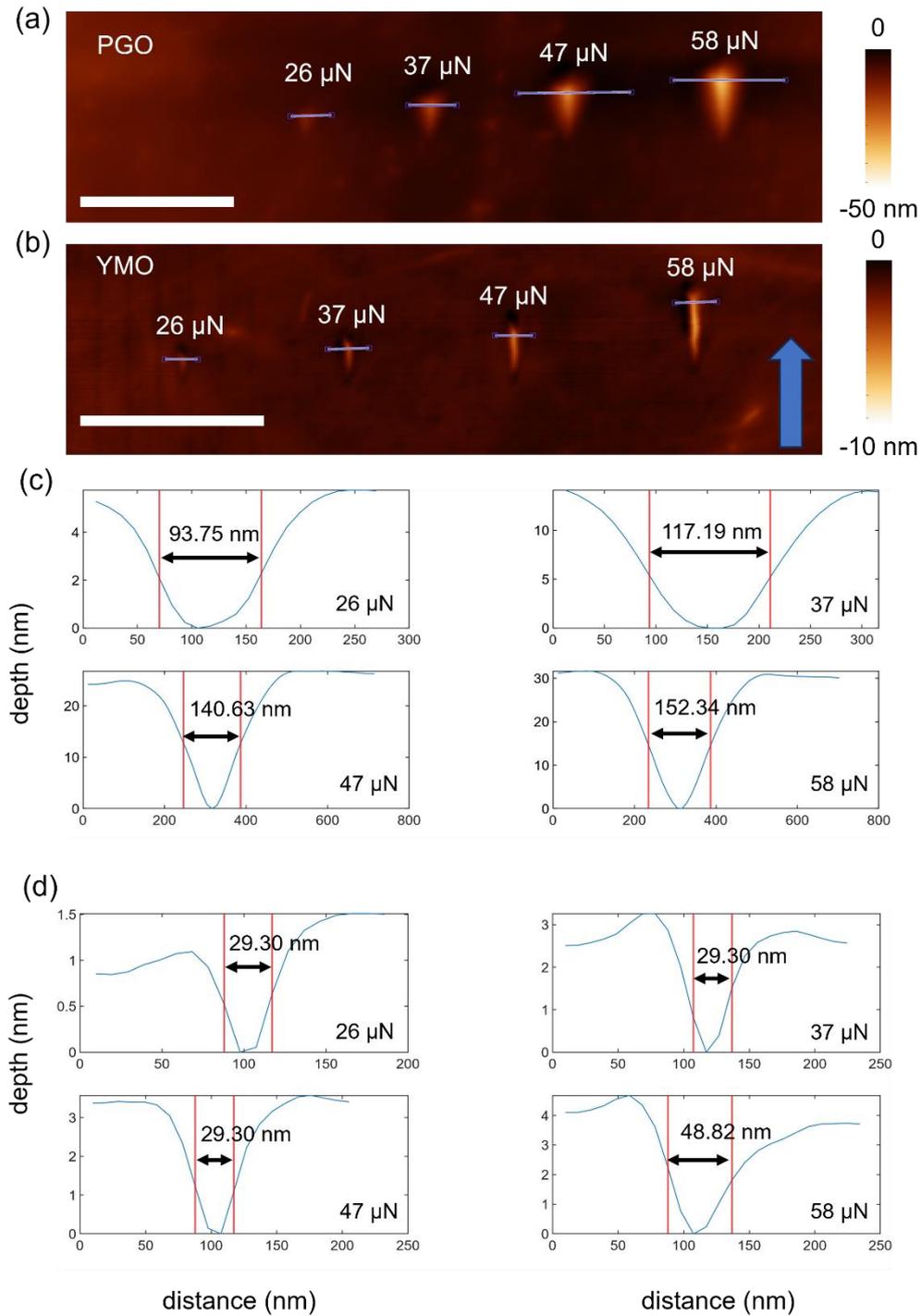

**Figure S1:** Hardness comparison of $Pb_5Ge_3O_{11}$ (PGO) and $YbMnO_3$ (YMO) through AFM-tip-based indentation. Topography of the indented region in PGO (a) and $YbMnO_3$ (b). The blue arrow represents the cantilever direction. (c) and (d) are the line profiles across the indented locations in PGO and YMO, respectively. An indented location is marked by the maximum force applied during the indentation. The line profiles are also shown in (a) and (b). Hardness(YMO) : Hardness(PGO) are found to be 10.24, 16.00, 23.04, and 9.73 at forces of 26 µN, 37 µN, 47 µN and 58 µN, respectively. The indentations are performed with an 80 N/m diamond tip. The scale bar for panels (a) and (b) is 0.92 µm.

2. **PFM and c-AFM maps collected from an area prior to milling**

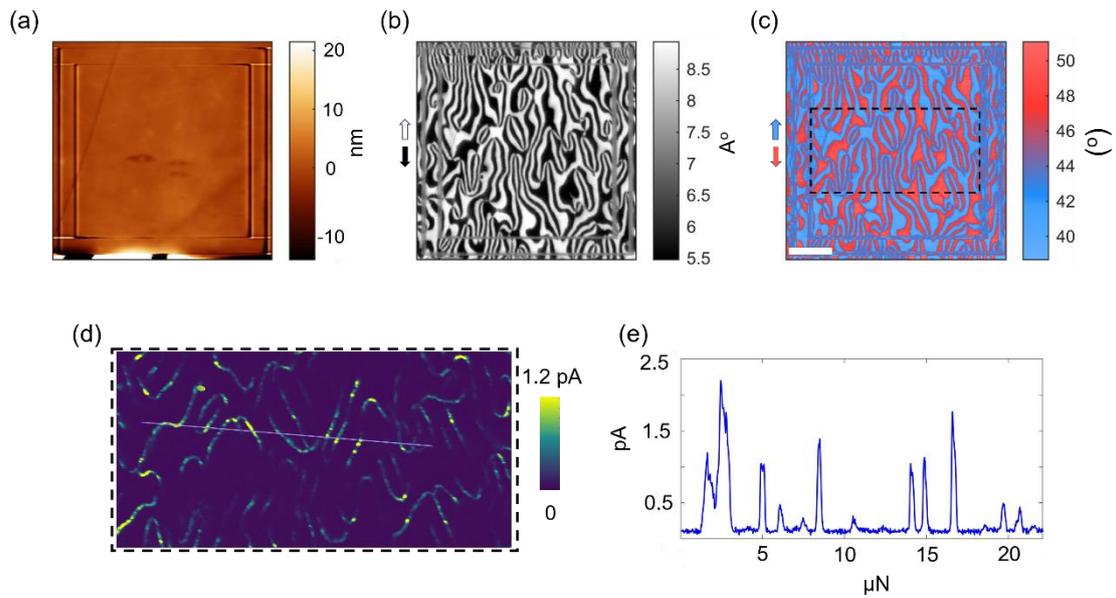

**Figure S2:** (a) Topography (b) PFM-Amplitude (c) PFM-phase collected on the same region in Figure 5 (main text) prior to milling. The scale bar is 8 µm. (d) c-AFM map of the highlighted area (black dotted box) in (c). (e) is a line profile taken along the marked line in (d) shown enhanced conduction only at the T-T domain walls. No clear difference in the conductance of the H-H domain walls and the domains can be seen. The PFM and c-AFM maps were collected using a 350 N/m diamond tip with a 10 V AC bias at a frequency of 4 MHz and a 6 V DC bias, respectively.

3. **Approach adopted to obtain the line profiles at different stages of AFM tomography**

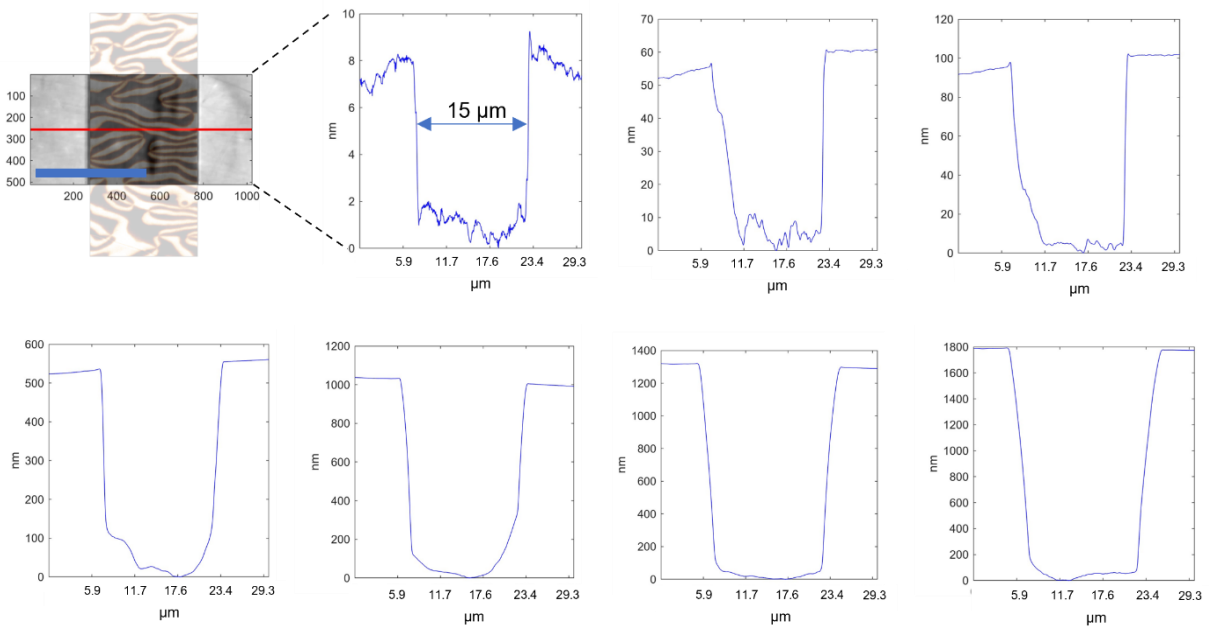

**Figure S3**: This figure illustrates the line profile across the milled area at different stages till a depth of 1.8 µm. A PFM map acquired during milling is superimposed on the topography map to show the direction in which the line profile is collected. The line profile is taken perpendicular to the length of the milled box. The depth was estimated by subtracting the maximum from the minimum value within a line profile drawn through the centre of the milled

box. Therefore, the milled depths referred to in the main text are the upper limits at each milling stage. The scale bar is 15 μm.

**4. Analysis for calculating charge states at domain walls from the PFM data**

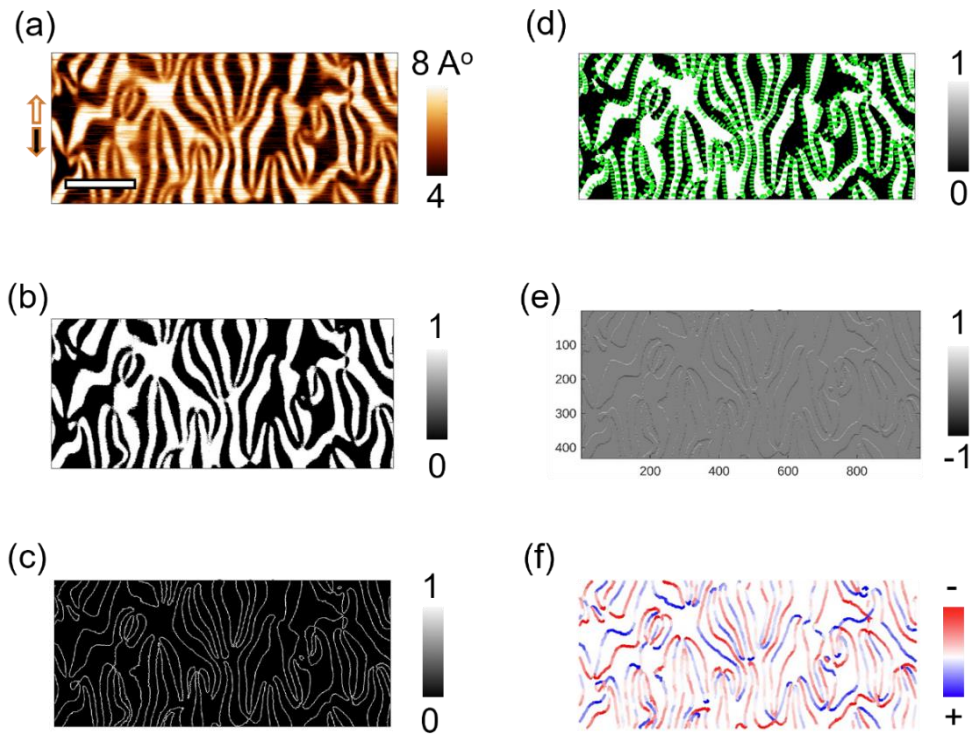

**Figure S4:** Image analysis for extracting charge states at the domain walls (a) PFM amplitude image at a depth of around 400 nm (b) Binarized PFM image (c) Location of domain walls (d) normal vectors calculated at each domain wall segment and overlaid on panel (b). (e) gradient map of (b) along the y-axis. (f) calculated charge state of each domain wall segment. The scale bar is 6 μm.

**5. Circuit modelling of conducting pathways**

A simulation of the model shown in **Figure S5** (a) was undertaken in MATLAB using Simulink and the voltage at the junction points relative to the tip was computed. The domain wall segments are treated as resistors of resistance R and the vertexes as current splitters. As demonstrated in **Figure S5** (b), we can anticipate a significant portion of the voltage to drop in the first 2 to 4 junctions if only the effect of vertexes was considered. Therefore, the fluctuations in domain wall orientation relative to the polarisation axis and its corresponding charge state distribution can be expected to have the biggest impact within the range of 2 to 4 vertex points from the material top surface. It is worth noting that this model presumes a limiting case where the current signal to be exclusively confined within the T-T domain walls, or the relative domain wall to domain conductivity is infinite, and the electric field profile due to the point (tip)-plane (ground) geometry is not considered. In a previous study, Roede *et al.*

took into account the relative conductivities of the domain walls compared to the domains and derived the voltage profile (relative to the tip) for a straight single domain wall[v]:

$$V(z) = V_o \left(1 - \mathrm{arctanh}\left[\sqrt{\frac{\sigma^*-1}{\sigma^*+1}}\frac{\sqrt{1+(^z/_r)^2}-1}{^z/_r}\right] \Big/ \mathrm{arctanh}\left[\sqrt{\frac{\sigma^*-1}{\sigma^*+1}}\right]\right)$$

$$\text{where } \sigma^* = \frac{\left(\frac{\sigma_{DW}}{\sigma_{bulk}}-1\right)\epsilon}{r}$$

Here, $V(z)$ is the voltage at a distance $z$ underneath the tip within a straight domain wall of width $\epsilon$, for a tip of contact radius $r$ at a potential $V_o$. For a domain wall of width 1 nm the predicted cutoff length (z at which $V(z) = V(z)/4$) at different tip radius and relative domain wall conductivities is plotted in **Figure S5**(c).

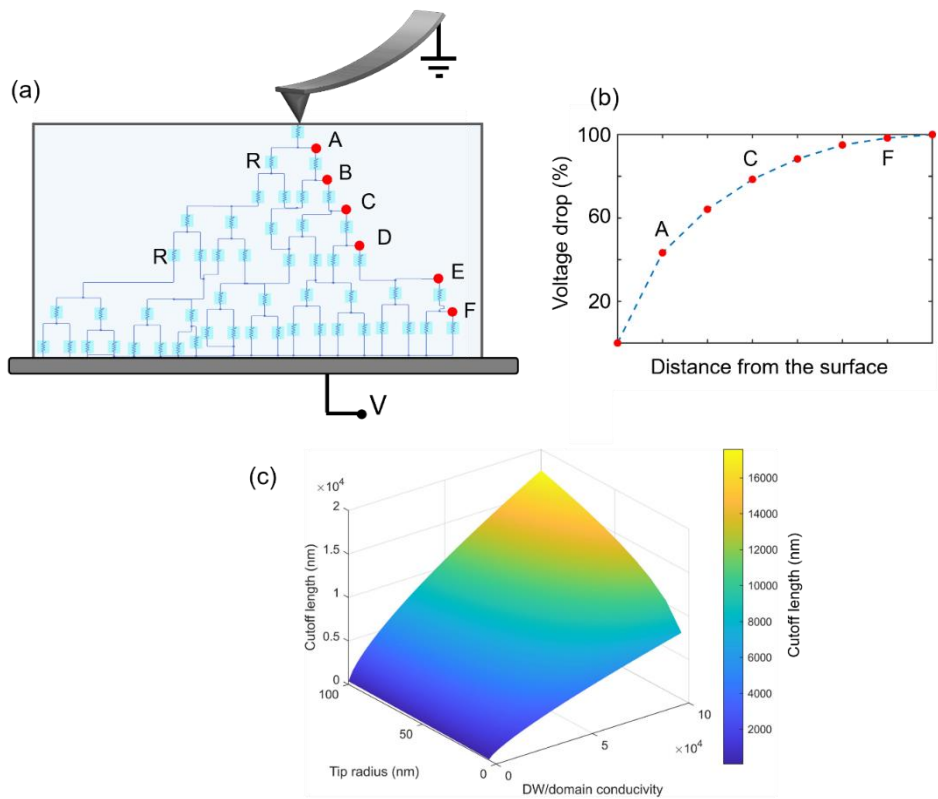

**Figure S5:** Resistor network model in hexagonal manganites: a) Schematic of a randomly generated resistor network, between the tip and the sample base in c-AFM setup. Points A, B, C, D, E, and F signify junction points at various depths from the sample surface. b) The plot shows the voltage profile in the domain wall network when we move from the tip to the base. (c) Estimated cutoff length (nm) at varied relative domain wall conductivities and tip radius (nm) for a 1nm wide straight domain wall.